# Shallow carrier traps in hydrothermal ZnO crystals

C. Ton-That*, L. L. C. Lem, and M. R. Phillips

*School of Physics and Advanced Materials, University of Technology Sydney,*

*P.O. Box 123, Broadway, NSW 2007, Australia*

F. Reisdorffer, J. Mevellec, and T.-P. Nguyen

*Institut des Matériaux Jean Rouxel, Universite de Nantes, 44322 Nantes Cedes 03,*

*France*

C. Nenstiel and A. Hoffmann

*Institut für Festkörperphysik, Technische Universität Berlin, Hardenbergstr. 36*

*10623 Berlin, Germany*

*\* Corresponding author: Cuong.Ton-That@uts.edu.au*

**Abstract**

Native and hydrogen-plasma induced shallow traps in hydrothermally grown ZnO crystals have been investigated by charge-based deep level transient spectroscopy (Q-DLTS), photoluminescence and cathodoluminescence microanalysis. The as-grown ZnO exhibits a trap state at 23 meV, while H-doped ZnO produced by plasma doping shows two levels at 22 meV and 11 meV below the conduction band. As-grown ZnO displays the expected thermal decay of bound excitons with increasing temperature from 7 K, while we observed an anomalous behaviour of the excitonic emission in H-doped ZnO, in which its intensity increases with increasing temperature in the range 140 – 300 K. Based on a multitude of optical results, a qualitative model is developed which explains the *Y* line structural defects, which act as an electron trap with an activation energy of 11 meV, being responsible for the anomalous temperature-dependent cathodoluminescence of H-doped ZnO.

# Contents

## 1. Introduction

Nominally undoped ZnO exhibits n-type conductivity due to the existence of shallow intrinsic defects and the high solubility of extrinsic donor-like impurities such as hydrogen [1, 2]. It has been established that hydrogen strongly affects the electronic properties of ZnO. Calculations based on density-functional theory (DFT) by Van de Walle *et al.* [3, 4] indicated that interstitial hydrogen ($H_i$) and hydrogen trapped at an oxygen vacancy ($H_O$) can act as two shallow donors, which can be a cause of n-type conductivity in ZnO. These theoretical results have been supported by experiments, which showed that the n-type conductivity and the near-band-edge (NBE) luminescence efficiency of ZnO can be enhanced at the expense of defect-related emissions via hydrogen incorporation [5-10]. Although the origins of defect-related emissions in ZnO are still not thoroughly understood, the quenching effect suggests that hydrogen interacts with native defects in some way. There have been reports of a stable hydrogen-related complex formed by zinc vacancy ($V_{Zn}$) acceptor and two H atoms [11]. As a reactive and common impurity, understanding the interaction of hydrogen with native defects in ZnO is not only of fundamental interest but also of technological importance.

Deep defect levels in ZnO have been previously investigated by the capacitance-based deep level transient spectroscopy (C-DLTS) technique as complementary investigations to optical studies [12-16]. Deep traps below the conduction band minimum such as $E_1$ (0.12 eV), $E_2$ (0.10 eV) and $E_3$ (0.29 – 0.30 eV) and $E_4$ (0.53 eV) were reported for ZnO grown by different methods; these traps were assigned to significant impurities or point defects in the bulk [12, 16-18]. Surface traps in ZnO have previously been investigated by C-DLTS [14]; however, C-DLTS signals can be severely distorted as the trapped charge can be varied as a function of applied bias. While the conventional C-DLTS is a powerful technique to investigate the behaviour of traps, it is not capable of identifying shallow trap states due to carrier freeze-out at low temperatures and immeasurably small capacitance of ZnO junctions [19, 20]. Electrically active shallow defects play a pivotal role in determining the electrical and optical properties of ZnO; however, controlling the behaviour of these defects remains a major challenge. Conversely, the charge-based deep level transient spectroscopy (Q-DLTS) used in this work has been specifically developed to facilitate probing of shallow trapping states and accordingly has been applied to the investigation of these defect centers in ZnO crystals. In contrast to C-DLTS, Q-DLTS is an isothermal technique in which the transient process of trapped charge (not capacitance) is measured after voltage stimulation. In this work, combined results from Q-DLTS and variable-temperature photoluminescence (PL) and cathodoluminescence (CL) spectroscopy allow quantitative evaluation of shallow level defects in the near surface region of the as-grown and H-doped ZnO crystals.

Traps in the near-surface region of a semiconductor can be investigated through the use of Q-DLTS because the sensitivity of charge transient detection is enhanced for surface states in metal-semiconductor junctions [21]. In the Q-DLTS method, cyclic bias pulses are applied to a Schottky barrier junction to excite carrier traps; the electron occupation of the traps is monitored by measuring the associated charge transients as the junction returns to thermal equilibrium. The charge transient is measured at two times ($t_1$ and $t_2$ from the beginning of discharge) and the charge $\Delta Q$ flowed through the circuit during the period $t_1 \to t_2$ is measured as a function of rate window $\tau = (t_2 - t_1)/\ln\left(\frac{t_2}{t_1}\right)$. For electrons, the gate timed charge difference is [22, 23]:

$$\Delta Q = Q(t_2) - Q(t_1) = Q_o[\exp(-e_n t_2) - \exp(-e_n t_1)] \quad [1]$$

where $Q_o$ is the total charge trapped during the filling pulse. The thermal emission rate $e_n$, according to Maxwell-Boltzmann statistics, can be expressed as [21, 22]:

$$e_n = \sigma \Gamma_n T^2 \exp\left(-\frac{E_a}{kT}\right) \quad [2]$$

where $\sigma$ is the capture cross section, $E_a$ the activation energy, and $\Gamma_n$ is a constant associated with the electron effective mass. The equations for holes are analogous. A common experimental approach is to keep the ratio $t_2/t_1 = \alpha$ constant; this leads to the function $\Delta Q(\tau)$ being maximum when the rate window is equal to the emission rate of the trap, i.e. $\tau_m^{-1} = \frac{\ln \alpha}{(\alpha-1)t_1} = e_n$. The activation energy of a trap and its cross section can therefore be obtained from an Arrhenius plot of equation [2], whereas the trap density can be calculated from the maximum $\Delta Q_{max}$ [24].

**2. Experimental details**

Identical samples from a single *a*-plane ZnO wafer (grown hydrothermally by the MTI Corp., USA, 0.5 mm thick) were used in this study. The crystal was polished both sides to 1 nm surface roughness. The samples of 5 mm × 5 mm square sections were cleaned in acetone and ethanol then rinsed in deionised water. One sample was exposed to radio-frequency plasma in hydrogen atmosphere (power 15 mW for 1 min with the sample kept at 200°C); this doping method leads to a near-surface hydrogen rich region as described in our previous work [25, 26]. No changes to the crystal structure were detected by Raman spectroscopy or X-ray diffraction. The Hall-effect characterization revealed that the as-grown and H-doped crystals have a carrier concentration of $4.4 \times 10^{15}$ to $7.1 \times 10^{17}$ cm$^{-3}$, respectively. The as-grown sample was also plasma cleaned by mild oxygen plasma to improve the reliability of electrical contacts according to the procedure as described by other workers [27]. A gold Schottky contact pad with a thickness of 150 nm was deposited by evaporation on one face of the crystal through a shadow mask, while In ohmic contact was established on the opposite face. Gold was used for the Schottky contact to prevent surface oxidation and to facilitate wire bonding. Current-voltage and isothermal Q-DLTS measurements were conducted in vacuum using the ASMEC-02 system, supplied by InOmTech Inc. Cathodoluminescence (CL) was performed using a FEI Quanta 200 SEM equipped with a liquid helium cold stage. The light emission was collected by a parabolic mirror and detected by a high-resolution Hamamatsu S7011-1007 CCD image sensor. Photoluminescence (PL) measurements were performed in a liquid helium bath cryostat at 7 K. The samples were excited by the 325 nm emission line of a HeCd laser and the emitted light was dispersed by a Spex-1404 double monochromator (spectral resolution 50 μeV) and detected by a bi-alkali photodetector.

## 3. Results and discussion

### 3.1. Shallow carrier traps investigated by Q-DLTS

A typical *I-V* characteristic of the ZnO/Au junction, showing a rectifying behaviour, is presented in Fig. 1. The forward bias was obtained with a positive bias applied to the Au electrode. Similar rectifying junctions on *c*-plane ZnO using Au electrodes have been reported previously [28]. The bulk resistivity of the crystal is 15.6 Ω·cm at room temperature. The sign of rectifying voltage (positive on the Au electrode) indicates the presence of *n*-type band bending with a depletion layer for electrons at the ZnO/Au interface. The relatively large reverse current, characteristic of Au Schottky contacts to ZnO, has been attributed to tunnelling conduction [14].

Figure 2(a) shows representative Q-DLTS traces from the as-grown crystal at temperatures ranging from 250 to 290 K. The Q-DLTS signal of the as-grown ZnO became negligible at temperatures below 250 K. As expected, the peak shifts gradually from a longer emission rate to a shorter emission rate with increasing temperature. The amount of charge emitted from the as-grown ZnO crystal increases with temperature, which is attributable to the addition of charge being trapped as the free carrier concentration increases with rising temperature. Figure 2(b) shows an Arrhenius plot derived from the data. Analysis of the Arrhenius plot using the effective mass $m_e^* = 0.24\ m_e$ [1] reveals a trap position of 23 ± 2 meV and a capture cross section, $\sigma = 2.6 \times 10^{-17}$ cm$^2$.

Selective Q-DLTS spectra recorded on the H-doped crystal at temperatures between 130 K and 290 K are presented in Figure 3(a), with the corresponding

Arrhenius plot shown in Figure 3(b). The Arrhenius plot shows the presence of two linear regions with different slopes, with a distinct break occurring at ~ 220 K (approximately the same temperature at which the Q-DLTS signal in the as-grown crystal completely disappears.) Two defect states can be determined from the slopes with activation energies of 22 ± 4 meV and 11 ± 2 meV and capture cross sections of $\sigma = 4.2 \times 10^{-17}$ and $3.8 \times 10^{-17}$ cm$^2$, respectively. To index Q-DLTS peaks, the subscript of E is used to indicate the energy level of a trap in meV. The difference in the activation energy of the H-doped $E_{22}$ and as-grown $E_{23}$ traps are within the experimental error of the Q-DLTS measurements. These two states, having similar activation energies and thermal emission rates, can be assigned to the same trap. We attribute the additional, lower energy state at 11 meV in the H-doped ZnO to a sub-surface state induced by the hydrogen plasma. While Q-DLTS itself cannot determine whether carriers in this trap are thermally activated to the valence band or conduction band, the coincidence of the H-related shallow state formation and the increase in electron density as revealed by the Hall effect measurement (from $4.4 \times 10^{15}$ to $7.1 \times 10^{17}$ cm$^{-3}$ upon hydrogen doping) confirms that a new donor state is responsible for the increased n-type conductivity. It was not possible to resolve the Q-DLTS peak of the H-doped crystal into components due to the small energy difference between the traps and inherent peak widths of the Q-DLTS spectra. Consequently, their trap energies could only be analysed using the Arrhenius plot. It is expected that the QDLTS signal in the H-doped ZnO at temperatures below 220 K arises predominantly from the 11 meV trap since the charge released from the $E_{22}$ trap is negligibly small over this temperature range [Fig 2(a)].

## 3.2. Luminescence properties

Figure 4(a) and (b) display representative CL spectra of the as-grown and H-doped ZnO crystals at various temperatures from 10 K to 300 K, together with the assignments for the spectral features. These spectra, collected with the same spectral resolution and the SEM operated at 15 kV corresponding to an excitation depth of 480 nm [25], display similar features in the near band edge (NBE) region. Notable spectral features are a bound exciton (BX) emission peak at 3.35 eV at 10 K and the phonon replicas of free exciton (FX) emission. The absence of the zero phonon FX emission in the NBE spectra is due to preferential binding of excitons to defects and impurities in ZnO. The strong sub-band gap FX-nLO signals in the CL spectra also suggest efficient self-absorption of FX photons in the crystal [25]. The phonon replica peaks (FX-1LO and FX-2LO) are separated by the longitudinal optical (LO) phonon energy $E_{LO} = 72$ eV [29]. This energy separation is insensitive to temperature over the measured temperature range but the well-resolved FX phonon replicas at low temperatures become thermally broadened with increasing temperature. The asymmetric shape of the NBE spectra with a long low-energy tail is due to the contribution from the higher-order phonon replicas. Comparison of the CL spectra of the as-grown and H-doped ZnO at 10 K reveals an increase in the BX intensity by approximately 50% after hydrogen incorporation, while the FX remains virtually unchanged for the CL spectra recorded at low temperatures ($< 60$ K), indicating the formation of additional recombination channels via neutral donor bound excitons.

The general trend displayed in Figure 4(a) and (b) is that the rapid thermal quenching of BX occurs with increasing temperature up to 150 K due to the delocalisation of bound excitons, while the FX replica intensities are largely unaffected. With a further increase in temperature, the BX emission is not detectable while the FX-1LO becomes the strongest feature in the spectral region. It is interesting to observe that the phonon-assisted FX recombination dominates the NBE emission in both the as-grown and H-doped ZnO at temperatures above 150 K. As the average kinetic energy of free excitons increases with rising temperature, the thermal redistribution leads to a smaller number of free excitons near the centre of the Brillouin zone, thus enhancing the participation of LO phonons in the radiative recombination. The strong LO-phonon-exciton coupling is consistent with the PL spectral evolution of ZnO with increasing temperatures [30, 31], with the coupling strength found to be four times greater than that of GaN [32]. While the overall features of the CL spectra are similar for the as-grown and H-doped ZnO at temperatures below 140 K, hydrogen has a striking effect on the FX and its phonon replicas: their intensities continues to rise in intensity with increasing temperature up to 320 K. This intriguing phenomenon is clearly seen in Figure 4(b), which shows the BX and FX-1LO peaks exhibiting an opposite dependence on temperature.

The enhancement of the NBE due to H doping is more evident in Fig 4(c), which shows the temperature dependence of the integrated intensity of the NBE emission for the samples. For both as-grown and H-doped ZnO, pronounced decreases of intensity with temperature over the range 10 – 140 K is typical of a CL spectrum that is dominated by shallow bound excitons. The activation energy for the luminescence quenching, determined from Arrhenius plots, are 11.6 and 6.5 meV for the as-grown and H-doped ZnO, respectively. The activation energy (11.6 meV) of the as-grown NBE is within the range of reported localisation energies for neutral donors bound excitons (10 to 28 meV) [1] and indicates that the thermal decay of bound excitons is the main mechanism for the thermal decay. On the other hand, the activation energy (6.5 eV) for the H-doped ZnO is significant smaller the localisation energies, indicating excited states of excitons are involved [33] (see resolved peaks in the high-resolution PL spectra below). The NBE emission of H-doped ZnO exhibits an anomalous behaviour, in which its intensity increases with increasing temperature over the range 140 – 300 K. The delocalisation and thermalisation of excitons cannot explain this peculiar temperature-dependent behaviour. At elevated temperatures (> 300 K), the CL quenching of H-doped ZnO was observed with an apparent activation energy of 18.7 meV, significantly smaller than the free-exciton binding energy (60 meV). This is not unexpected given the fact that the de-trapping of carriers from traps occurs over this temperature range (as demonstrated by Q-DLTS) and carriers might take part in thermally activated non-radiative recombination processes.

To gain further insight into the enhanced optical and electrical effects in the H-doped ZnO, high-resolution PL was conducted. Figure 5(a) shows low temperature (7 K) PL spectra of the as-grown and H-doped ZnO. Within the region of 3.14 – 3.34 eV, the notable spectral change after hydrogen plasma is the considerable enhancement of the peak at 3.336 eV, attributed to *Y* line [34, 35]. The *Y* line could not be resolved in the CL spectra, possibly due to the fact that the analysis depth (the CL generation depth is 490 nm at 15 kV) is significantly greater than the depth profile of surface extended defects induced by hydrogen plasma. The phonon replicas and two-electron satellite (TES) transition of the BX and FX are also observed in this spectral range. In an enlargement of the BX region (inset of Figure 5), the most intense lines in the as-grown ZnO are $I_4$ and $I_6$, which have been attributed to H and Al shallow donors [1]. Neutral-donor-bound exciton transition $I_9$ at 3.357 eV, which was recently verified by time-of-flight PL [36], appears as a shoulder with intensity about an order of magnitude weaker than those of $I_4$ and $I_6$. The presence of these impurities in the as-grown crystal is unsurprising since they are commonly present during the hydrothermal growth. For H-doped ZnO, the dominant peak at 3.362 eV (labelled $I_{4a}$) has previously attributed to an excited state of the H-related neutral donor bound exciton $I_4$ [33, 37]. Analysis of the BX spectral region reveals that $I_{4a}$ intensity increases in comparison with that of $I_4$ after hydrogen incorporation suggests that hydrogen dopants introduced by the plasma is in a different chemical state to those incorporated during the hydrothermal growth. Figure 5(b) displays the temperature-resolved PL spectra of the *Y* line and TES $I_4$ emission in the range of 7 – 75 K. The *Y* line emission cannot be resolved at temperatures above 75 K due to its overlap with the dominant $D^oX$ peak. Analysis of the thermalisation behaviour of the *Y* line emission yields a thermal activation energy of 11 meV. Although the absolute energy of the *Y* line is slightly shifted compared with those in Cermet and EaglePicher ZnO samples [35], possibly due to strain, this energy is in excellent agreement with previously reported activation values of 10 – 12 meV for the *Y* line [34, 35]. The *Y* line recombination has been found in ion implanted ZnO crystals and attributed to extended structural defects [35]; our results here indicate that similar structural defects could be introduced in ZnO by hydrogen plasma.

## 4. Discussion

The dominant electron trap $E_{11}$ in the H-doped ZnO could be associated with either hydrogen dopants or *Y*-line structural defects induced by the plasma process. However, the shallow $E_{11}$ trap is unlikely to be a H-induced donor state, which has a much larger ionisation energy ($E_i$(H) = 47 − 53 meV [38, 39]). The nature of the shallow $E_{11}$ strap is currently undetermined in the literature, with its activation energy being significantly smaller than that of the shallowest traps detected previously by C-DLTS in hydrothermal ZnO (55 meV) [40] or in ZnO thin films deposited by pulsed laser deposition (31 meV) [41]. The trap depth of $E_{11}$ is consistent with the thermal activation energy of the *Y* line (see section 3.2 above). The *Y*-line luminescence exhibits an unusual behaviour with increasing temperature - despite possessing a large localisation energy of 38 meV, the thermal activation energy of the *Y* line is only 11 meV. Such thermal behaviour of the *Y*-line can be well explained if its thermal activation energy corresponds to the detachment of an electron from the $E_{11}$ shallow trap, rather than the detachment of an entire exciton, as suggested by Wagner *et al*. [35]. Furthermore, the Q-DLTS peak intensity of the $E_{11}$ trap does not saturate with increasing pulse-width in the range of 0.1 to 5 ms, which is highly characteristic of traps associated with extended defects that can trap multiple charges [42]. This result further supports the contention that the $E_{11}$ trap is associated with the *Y* line structural defects.

In Q-DLTS, the maximum depth of the detectable trap volume is the width of the depletion layer [19]. From the standard calculation of the depletion layer at the applied bias, we can estimate that the $E_{11}$ trap arises from depths up to 60 nm. Based on the above discussion, the $E_{11}$ trap is therefore be attributable to *Y* line structural defects in the near-surface region that are induced by the plasma process. Our previous depth-resolved CL microanalysis showed the hydrogen plasma causes the greatest changes to the defect structure in the top 1.5 μm layer of the crystal [25]. A possible model that embodies the observed temperature-dependent luminescence and defect properties could be based on the release of electrons from *Y*-line structural defects at temperatures above 140 K, leading to the generation of additional excitons. At temperatures below 140 K, the *Y*-line defects act like electrically active shallow traps with an electron binding energy of 11 meV. Furthermore, the fact that the activation energy of extended defects is significantly smaller than the localisation energy of the *Y*-line emission suggests that *Y* line excitons do not bind to the trap as a whole quasiparticle, but rather as a separate electron and hole bound by weak Coulombic interaction.

The origin of the 23 meV electron trap in the as-grown ZnO is unknown at present but hydrogen and aluminium are prime candidates because their emission lines are most clearly detected by PL at low temperatures. A shallow donor level at 20 meV in hydrothermally grown ZnO, close to the $E_{23}$ trap, has been previously reported by other workers using thermal admittance spectroscopy (TAS) [40], but this trap state was not detectable by C-DLTS, probably due to carrier freeze-out at low temperatures. The slight increase in the cross section of the $E_{23}$ trap after the hydrogen plasma might be due to reduction in the number of competitive traps as the ZnO surface is passivated by hydrogen.

**5. Summary**

In summary, Q-DLTS has been applied successfully to ZnO crystals to reveal shallow traps at 11 and 23 meV below the conduction band edge, which have not been previously detected by C-DLTS. Hydrogen doping by plasma causes significant enhancement of the *Y*-line emission at 3.336 eV and the anomalous behaviour of the excitonic emission in H-doped ZnO over the temperature range 140 – 300 K. Based on the presented experimental results, the unusual temperature-dependent luminescence of H-doped ZnO is explained by a model in which generation of additional excitons occurs as a result of electron release from the 11 meV trap.

**References**

[1] B. K. Meyer, H. Alves, D. M. Hofmann, W. Kriegseis, D. Forster, F. Bertram*, et al.*, Phys. Status Solidi B-Basic Res. **241**, 231 (2004).

[2] A. Janotti and C. G. Van de Walle, Rep. Prog. Phys. **72**, 126501 (2009).

[3] A. Janotti and C. G. Van de Walle, Nat. Mater. **6**, 44 (2007).

[4] C. G. Van de Walle, Phys. Rev. Lett. **85**, 1012 (2000).

[5] P. F. Cai, J. B. You, X. W. Zhang, J. J. Dong, X. L. Yang, Z. G. Yin*, et al.*, J. Appl. Phys. **105**, 083713 (2009).

[6] Z. Zhou, K. Kato, T. Komaki, M. Yoshino, H. Yukawa, M. Morinaga*, et al.*, J. European Ceram. Soc. **24**, 139 (2004).

[7] T. Sekiguchi, N. Ohashi, and Y. Terada, Jpn. J. Appl. Phys. Part 2 - Lett. **36**, L289 (1997).

[8] C. C. Lin, H. P. Chen, H. C. Liao, and S. Y. Chen, Appl. Phys. Lett. **86**, 183103 (2005).

[9] A. Dev, R. Niepelt, J. P. Richters, C. Ronning, and T. Voss, Nanotechnology **21**, 065709 (2010).

[10] C. Ton-That, L. Weston, and M. R. Phillips, Phys. Rev. B **86**, 115205 (2012).

[11] E. V. Lavrov, J. Weber, F. Borrnert, C. G. Van de Walle, and R. Helbig, Phys. Rev. B **66**, 165205 (2002).

[12] F. D. Auret, S. A. Goodman, M. J. Legodi, W. E. Meyer, and D. C. Look, Appl. Phys. Lett. **80**, 1340 (2002).

*New Journal of Physics*

[13]   W. Mtangi, F. D. Auret, W. E. Meyer, M. J. Legodi, P. J. J. van Rensburg, S. M. M. Coelho, *et al.*, J. Appl. Phys. **111**, 094504 (2012).

[14]   Z. Q. Fang, B. Claflin, D. C. Look, Y. F. Dong, H. L. Mosbacker, and L. J. Brillson, J. Appl. Phys. **104**, 063707 (2008).

[15]   M. Grundmann, H. von Wenckstern, R. Pickenhain, T. Nobis, A. Rahm, and M. Lorenz, Superlattices Microstruct. **38**, 317 (2005).

[16]   R. Heinhold, H. S. Kim, F. Schmidt, H. von Wenckstern, M. Grundmann, R. J. Mendelsberg, *et al.*, Appl. Phys. Lett. **101**, 062105 (2012).

[17]   G. Brauer, W. Anwand, W. Skorupa, J. Kuriplach, O. Melikhova, C. Moisson, *et al.*, Phys. Rev. B **74**, 045208 (2006).

[18]   L. Vines, J. Wong-Leung, C. Jagadish, E. V. Monakhov, and B. G. Svensson, Physica B **407**, 1481 (2012).

[19]   V. I. Polyakov, A. I. Rukovishnikov, N. M. Rossukanyi, and V. G. Ralchenko, Diam. Relat. Mat. **10**, 593 (2001).

[20]   F. Schmidt, S. Muller, H. von Wenckstern, C. P. Dietrich, R. Heinhold, H. S. Kim, *et al.*, Appl. Phys. Lett. **103** (2013).

[21]   I. Thurzo, R. Beyer, and D. R. T. Zahn, Semicond. Sci. Technol. **15**, 378 (2000).

[22]   D. V. Lang, J. Appl. Phys. **45**, 3023 (1974).

[23]   B. M. Arora, S. Chakravarty, S. Subramanian, V. I. Polyakov, M. G. Ermakov, O. N. Ermakova, *et al.*, J. Appl. Phys. **73**, 1802 (1993).

[24]   O. Gaudin, R. B. Jackman, T. P. Nguyen, and P. Le Rendu, J. Appl. Phys. **90**, 4196 (2001).

[25]   L. C. L. Lem, C. Ton-That, and M. R. Phillips, J. Mater. Res. **26**, 2912 (2011).

[26]   L. Weston, C. Ton-That, and M. R. Phillips, J. Mater. Res. **27**, 2220 (2012).

[27] Y. F. Dong, Z. Q. Fang, D. C. Look, D. R. Doutt, G. Cantwell, J. Zhang, *et al.*, J. Appl. Phys. **108**, 103718 (2010).

[28] L. J. Brillson and Y. C. Lu, J. Appl. Phys. **109**, 121301 (2011).

[29] K. Maeda, M. Sato, I. Niikura, and T. Fukuda, Semicond. Sci. Technol. **20**, S49 (2005).

[30] C. Sturm, H. Hilmer, R. Schmidt-Grund, and M. Grundmann, New J. Phys. **11**, 073044 (2009).

[31] W. Shan, W. Walukiewicz, J. W. Ager, K. M. Yu, H. B. Yuan, H. P. Xin, *et al.*, Appl. Phys. Lett. **86**, 191911 (2005).

[32] T. Makino, C. H. Chia, N. T. Tuan, Y. Segawa, M. Kawasaki, A. Ohtomo, *et al.*, Appl. Phys. Lett. **76**, 3549 (2000).

[33] B. K. Meyer, in *From Fundamental Properties Towards Novel Applications*, edited by F. Klingshirn, W. Andreas, A. Hoffmann and J. Geurts (Springer, New York, 2010), Chap. 7.

[34] H. Alves, D. Pfisterer, A. Zeuner, T. Riemann, J. Christen, D. M. Hofmann, *et al.*, Opt. Mater. **23**, 33 (2003).

[35] M. R. Wagner, G. Callsen, J. S. Reparaz, J. H. Schulze, R. Kirste, M. Cobet, *et al.*, Phys. Rev. B **84**, 035313 (2011).

[36] T. V. Shubina, M. M. Glazov, N. A. Gippius, A. A. Toropov, D. Lagarde, P. Disseix, *et al.*, Phys. Rev. B **84**, 075202 (2011).

[37] B. K. Meyer, J. Sann, S. Eisermann, S. Lautenschlaeger, M. R. Wagner, M. Kaiser, *et al.*, Phys. Rev. B **82**, 115207 (2010).

[38] E. V. Lavrov, F. Herklotz, and J. Weber, Phys. Rev. B **79**, 165210 (2009).

[39] G. A. Shi, M. Stavola, S. J. Pearton, M. Thieme, E. V. Lavrov, and J. Weber, Phys. Rev. B **72**, 195211 (2005).

[40]    L. Vines, E. V. Monakhov, R. Schifano, W. Mtangi, F. D. Auret, and B. G. Svensson, J. Appl. Phys. **107**, 103707 (2010).

[41]    F. D. Auret, W. E. Meyer, P. J. J. van Rensburg, M. Hayes, J. M. Nel, H. von Wenckstern*, et al.*, in *Proceedings of the 17th International Vacuum Congress/13th International Conference on Surface Science/International Conference on Nanoscience and Technology*, edited by L. S. O. Johansson, J. N. Andersen, M. Gothelid, U. Helmersson, L. Montelius, M. Rubel, J. Setina and L. E. Wernersson (Iop Publishing Ltd, Bristol, 2008), Vol. 100.

[42]    D. C. Look, Z. Q. Fang, S. Soloviev, T. S. Sudarshan, and J. J. Boeckl, Phys. Rev. B **69**, 195205 (2004).

# Shallow carrier traps in hydrothermal ZnO crystals

**Figure Captions**

**Fig. 1.** IV characteristics of Au Schottky contacts deposited on the H-doped ZnO crystal, measured at 300 K.

**Fig. 2.** (a) Q-DLTS spectra obtained from the as-grown ZnO in the temperature range 250–300 K using charging voltage $\Delta V = 1$ V and charging time $t_C = 1$ s. (b) Arrhenius plot derived from the data in (a) indicating a state with an average activation energy of $23 \pm 2$ meV.

**Fig. 3** (a) Q-DLTS spectra obtained from H-doped ZnO using charging voltage $\Delta V = 1$ V and charging time $t_C = 1$ s. The spectra shown were recorded within the temperature range 130–290 K. (b) Arrhenius plot derived from the data in (a) indicating two states with energies of $22 \pm 4$ meV and $11 \pm 2$ meV. Hydrogen doping generates a new defect state, which becomes activated at ~ 150 K.

**Fig. 4.** Evolution of CL spectra in the NBE region with temperature for (a) as-grown ZnO and (b) H-doped ZnO crystals (e-beam parameters $E_B = 15$ keV, $I_B = 5.7$ nA). The LO-phonon replicas of FX and BX transitions are indicated by arrows. With increasing temperature the BX transition turns into the FX transition. (c) Integrated intensities of the NBE emission plotted as a function of temperature for the as-grown and H-doped ZnO. Solid curves are fits to the data providing the activation energies for the luminescence quenching, which are distinctly dependent on sample type and temperature range.

**Fig. 5.** (a) High resolution PL spectra taken at 7 K for the as-grown and H-doped ZnO crystals, showing the $Y$ line and phonon replicas of FX and $D^oX$. Several I lines in the $D^oX$ region after normalisation is displayed in the inset. The enhancement of I lines, especially the excited states $I_{4a}$, results in the broadening of the $D^oX$ peak. (b) PL spectra of the *Y* line and TES $I_4$ emissions for the H-doped ZnO as a function of temperature between 7 and 75 K.

Figure 1

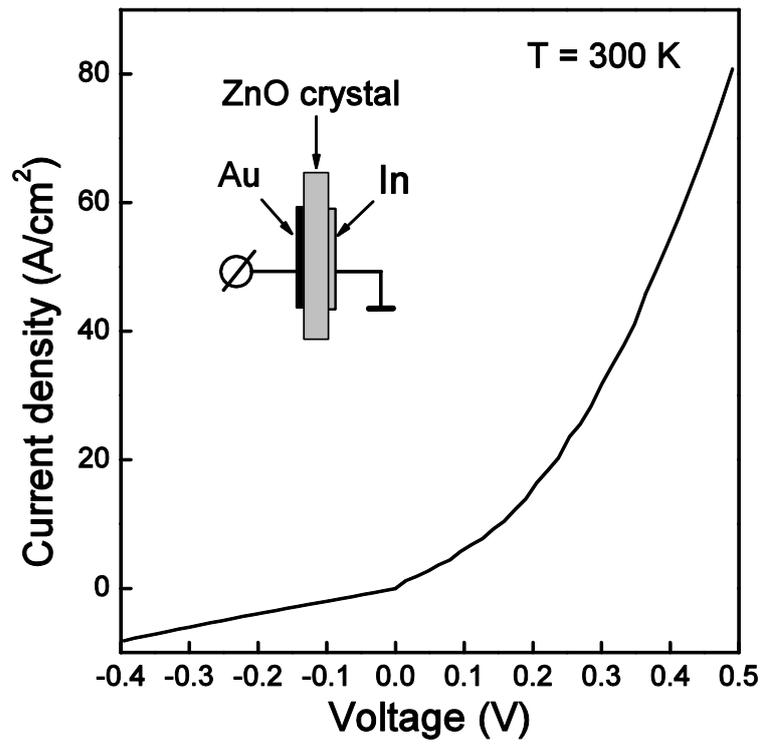

Figure 2
(a)

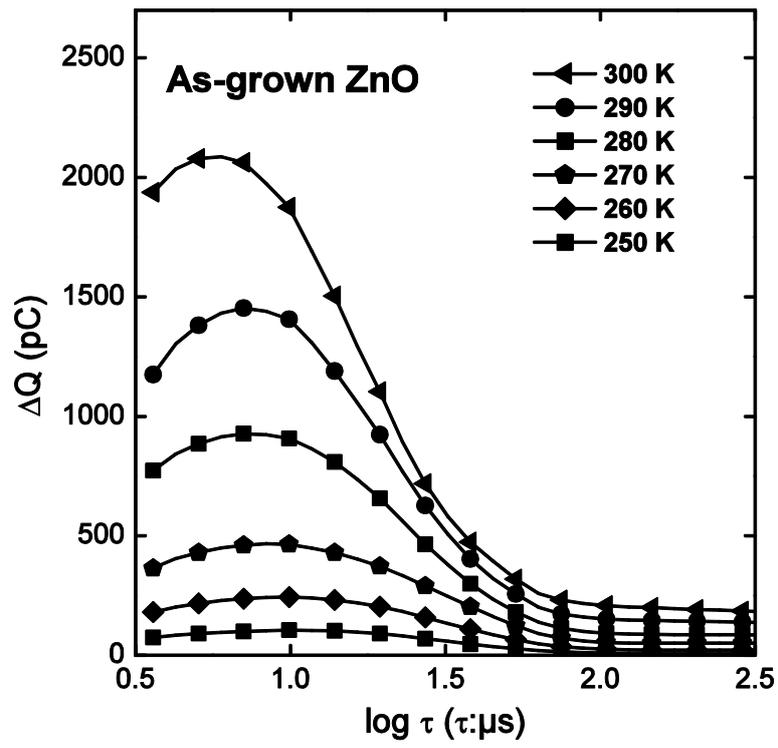

(b)

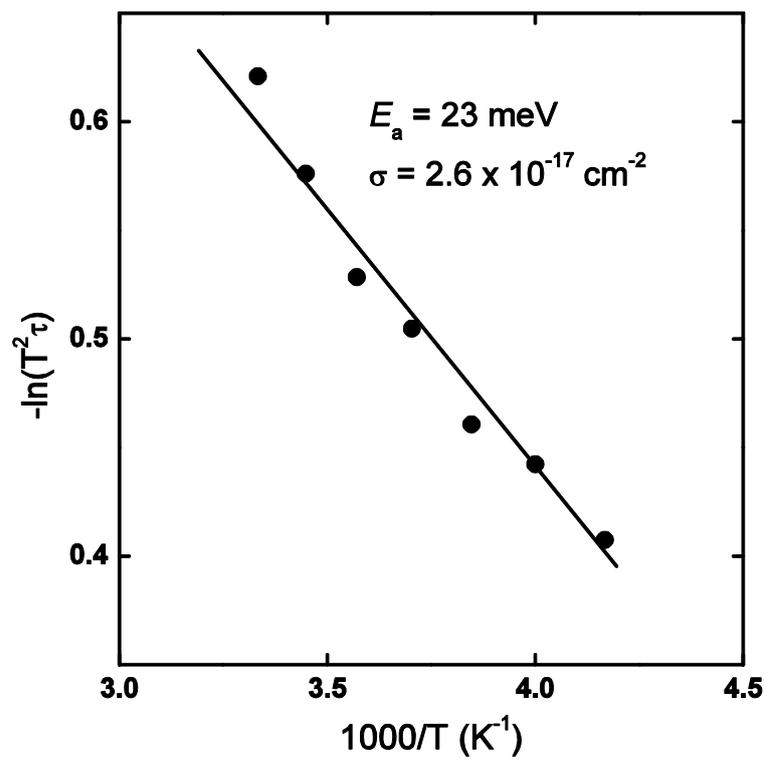

Figure 3
(a)

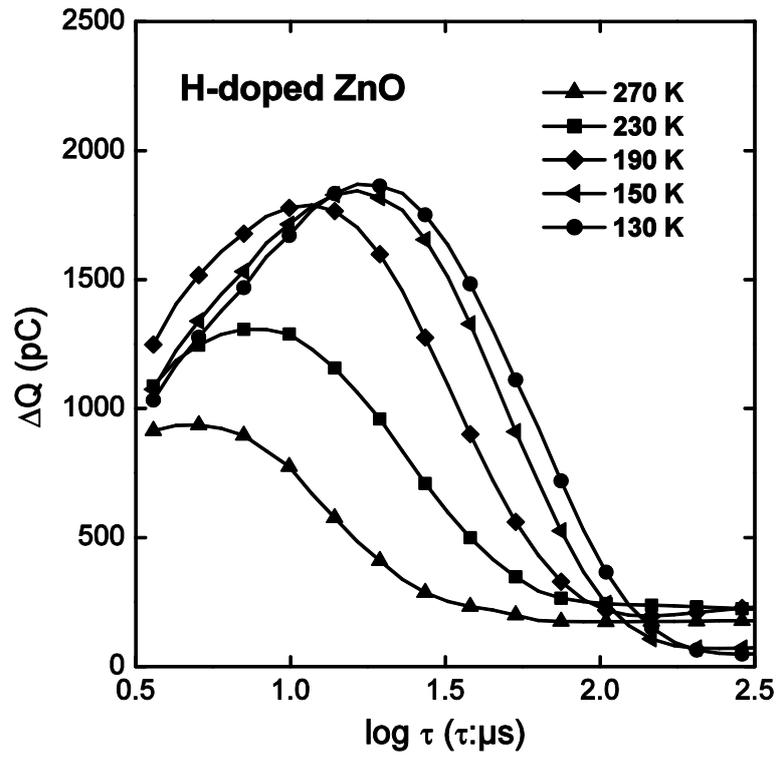

(b)

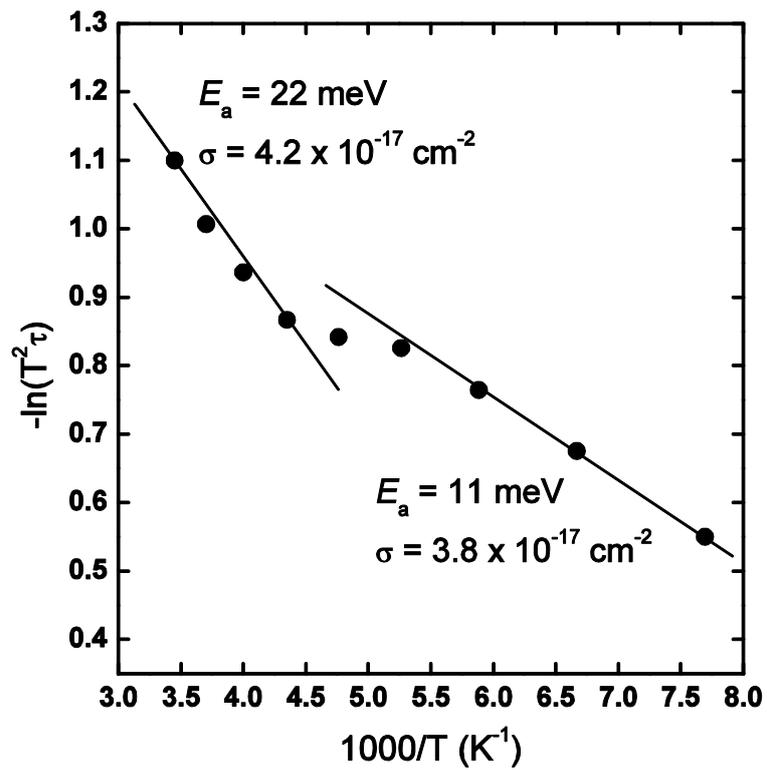

Figure 4

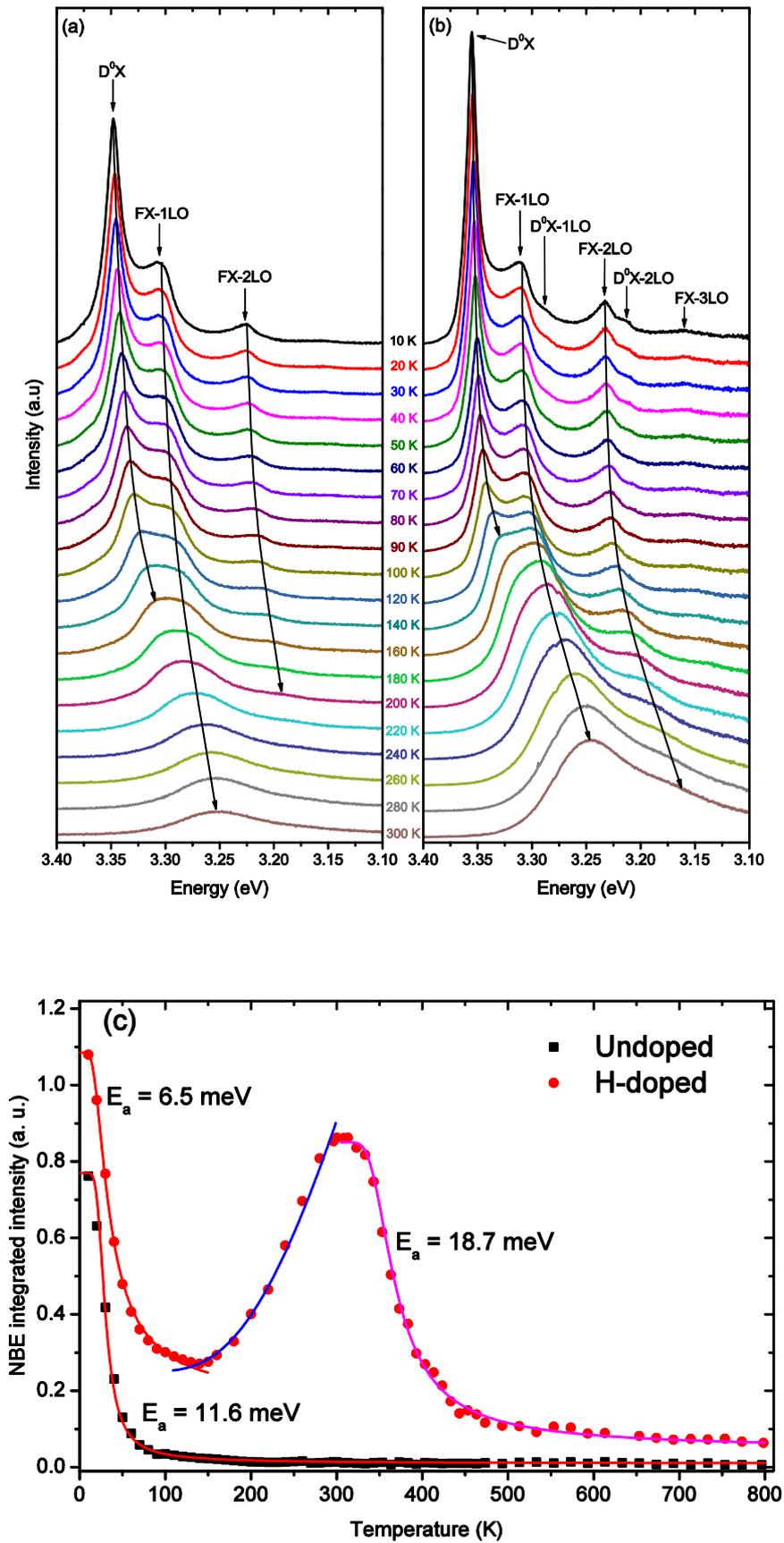

Figure 5

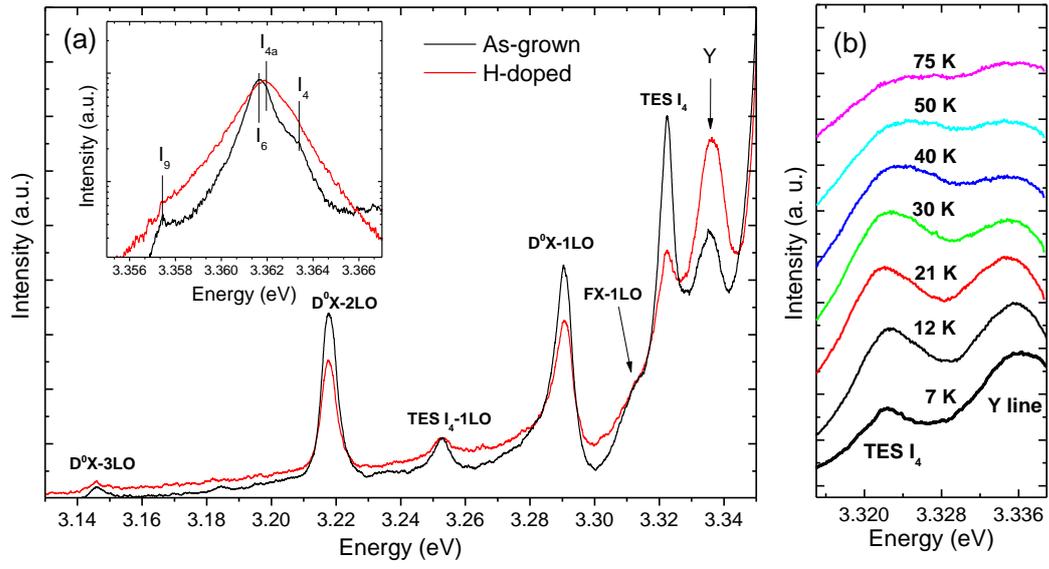